\documentstyle[12pt]{article}

\def\be{\begin{equation}}
\def\ee{\end{equation}}
\def\a{\alpha}
\def\e{\epsilon}
\def\ra{\rangle}
\def\la{\langle}
\def\d{\partial}

\begin{document}

\begin{center}
{\bf Jain hierarchy for the Fractional Quantum Hall Effect.}
\end{center}
\vspace{0.2in}
\begin{center}
{\large A.A.Ovchinnikov}
\end{center}

\begin{center}
{\it Institute for Nuclear Research, RAS, Moscow}
\end{center}

\vspace{0.1in}

\begin{abstract}

We propose the effective hierarchical partition function which is 
able to describe both the Jain states and the Jain-type hierarchical 
states. Using this partition function (effective Lagrangian) we calculate 
the charge of the quasiparticle excitations. We show that the Jain-type 
hierarchical states are equivalent to the system of anyons in the 
external magnetic field. 

\end{abstract}

\vspace{0.2in}

{\bf 1. Introduction}

\vspace{0.2in}

At present time it is well known that almost all of the fractional 
quantum Hall effect (FQHE) states observed experimentally (for example 
see the recent papers \cite{Pan}) are the members of the Jain sequence 
\cite{J} of FQH states. The rest of the states are the so called 
hierarchical states which is nothing else but the FQH states of 
quasiparticles i.e. the Jain-type eigenstates of quasiparticles 
(or composite fermions). 
Theoretically although the corresponding filling fractions are known 
\cite{LF03} their derivation remains somewhat obscure. 
It is the goal of the present letter to make these questions more 
clear. In particular we show that the effective hierarchical Lagrangian 
is nothing else but the Lagrangian of the system of anyons in the 
external magnetic field. 
Although quasiparticle (vortex) states for the Jain states cannot be 
constructed explicitly its quantum numbers and the external magnetic 
field can be evaluated. Using these quantities the hierarchical FQHE 
filling fractions can be predicted. 
To calculate them we propose the effective hierarchical partition 
function (effective Lagrangian) which is able to describe the Jain states 
and the Jain-type hierarchical states. 
In particular we calculate the charge of quasiparticles for the states 
of the Jain sequence. Although the value of this charge is known 
(for example see \cite{LF98}) its derivation remains obscure. 
Another goal is to present the simple derivation of the formula for 
the effective Lagrangian of fermions in the external magnetic field 
\cite{RSS}. Finally the third goal is to clarify the simple picture 
of anyons in the external magnetic field behind the hierarchical 
Jain-type FQHE states and calculate the conductivity by means of 
filling the integer number of Landau levels in the effective 
magnetic field. The FQHE is then follows from the non-zero contribution 
of the Chern-Simons term to the effective Lagrangian \cite{O},\cite{LF}. 
We believe all these points are an interesting contribution to the 
theory of FQHE.

\vspace{0.2in}

{\bf 2. Jain-type hierarchy for the Jain FQHE states.} 

\vspace{0.2in}

The object to be studied is the partition function corresponding to the 
eigenstate which consists of the stable FQHE parent state of the Jain 
sequence and the number of quasiparticles (Vortices) on the top of the 
parent state.  
The Jain sequence \cite{J} can be obtained in the framework of the 
Fermionic second-quantized approach \cite{O},\cite{LF}. The starting 
point of this approach is the anyonic transformation of the 
electron wave function of the form: 
\be
\Psi( {\bf x}_1,\ldots,{\bf x}_N)=\prod_{i<j}
\left(\frac{z_i-z_j}{|z_i-z_j|}\right)^{2k}
\Psi^{\prime}({\bf x}_1,\ldots,{\bf x}_N), 
\label{2k}
\ee
where $z_i=x_i+iy_i$ and $k$ is an integer. After that in the 
second-quantized language the Lagrangian takes the form: 
\be 
L=i\psi^{+}(\partial_0-ia_0)\psi+\psi^{+}(\partial-i{\bf a}-i{\bf A}_1)^2\psi 
+\frac{1}{2k}\frac{1}{4\pi}\e^{\mu\nu\a}a_{\mu}\d_{\nu}a_\a, 
\label{L}
\ee
where $\psi$- is the fermionic field and the effective magnetic field 
$H_1=H/(2nk+1)$ for the filling factor $\nu=\nu_0+\nu_1=n/(2nk+1)+\nu_1$. 
Equation (\ref{L}) is the starting 
point of our derivation of the FQHE hierarchical partition function. 
Roughly speaking we write $\psi\simeq\psi_0+V$, where the field $\psi_0$ 
corresponds to the Fermions in the parent Jain state and the field $V$ 
corresponds to the quasiparticles moving in the medium of the parent state. 
Clearly the resulting effective Lagrangian takes the following form: 
\[
L_{eff}=i\psi^{+}_0(\partial_0-ia_0-iA_0)\psi_0+
\psi^{+}_0(\partial-i{\bf a}-i{\bf A}-i{\bf A}_1)^2\psi_0+
\frac{1}{2k}\frac{1}{4\pi}\e^{\mu\nu\a}a_{\mu}\d_{\nu}a_\a+ 
\]
\be
iV^{+}(\partial_0-ia_0-iA_0)V+
V^{+}(\partial-i{\bf a}-i{\bf A}-i{\bf A}_1)^2V, 
\label{eff}
\ee
where the external effective magnetic field $H_1=H/(2nk+1)$ is the same for 
both type of Fermions $\psi_0$ and $V$. 
The partition function for the Lagrangian (\ref{eff}) is constructed 
with the help of the factors which fix the number of paricles (for 
example see eq.(\ref{psi}) below). Thus we get the effective partition 
function for the Jain-type hierarchical states for the states of the 
Jain sequence. This is the main result of the present letter.  
To proceed further we should integrate out the Fermi- field $\psi_0$. 
This procedure was performed in the number of papers \cite{RSS},\cite{LF}.  
Here we propose the simple way to obtain the effective action for the 
gauge fields. In fact one can see that the following equation is valid: 
\be
\int D\psi_0^{+}D\psi_0
\delta(\int d{\bf x}\psi_0^{+}\psi_0-\rho_0-\frac{1}{2\pi}nh)
e^{i\psi_0^{+}\d_0\psi_0+a_0(\psi_0^{+}\psi_0-\rho_0)+ 
\psi_0^{+}(\d-i{\bf a}-i{\bf A}_1)^2\psi_0}=
\label{psi}
\ee
\[
Z\exp(in\int dx \frac{1}{4\pi}\e^{\mu\nu\a}a_{\mu}\d_{\nu}a_\a+\ldots),  
\]
where the dots stand for the terms with two derivatives. 
Here the argument of the $\delta$- function which is introduced to fix 
the number of particles is chosen in such a way that at arbitrary $a_\mu$ 
(including non-zero value of the magnetic field $h=\e_{ij}\d_{i}a_{j}$) 
the number of the completely filled Landau levels is equal to $n$. 
Differentiating both sides of eq.(\ref{psi}) over $a_0$ we see that 
the result is correct which means that eq.(\ref{psi}) is correct. 
So we have the simple way to derive the equation (\ref{psi}).  
The same result could be obtained by means of fixing the number of particles 
with the help of the chemical potential. 
Thus according to eq.(\ref{psi}) 
after integration out of the Fermi- field $\psi_0$ we obtain the 
following effective Lagrangian depending on the statistical gauge field $a_\mu$ 
and the Vortex Fermi- field $V$: 
\be
L(a,A)=n\frac{1}{4\pi}\e^{\mu\nu\a}(a+A)_{\mu}\d_{\nu}(a+A)_\a+ 
\frac{1}{2k}\frac{1}{4\pi}\e^{\mu\nu\a}a_{\mu}\d_{\nu}a_\a+
\label{L1}
\ee
\[
V^{+}(\d-i{\bf a}-i{\bf A}-i{\bf A}_1)^2V+ (a_0+A_0)V^{+}V+A_{0}\rho_0,  
\]
where $A_\mu$ is the external electromagnetic field used to probe the system. 
Here the number of the completely filled Landau levels for the effective 
magnetic field after the transformation (\ref{2k}) equals $n$. 
Since the effective magnetic field $H_1=H/(2nk+1)$ the density $\rho_o$ 
corresponding to the field $\psi_0$ corresponds to the filling factor equal 
to the Jain sequence $\nu_0=n/(2nk+1)$. 
If the quasiparticles are absent at $n=1$ the Lagrangian (\ref{L1}) coincides 
with the dual effective Lagrangian for the Laughlin $1/m$ states proposed 
by Wen \cite{Wen}. 
To represent the field $V$ as anyons we introduce the new statistical gauge field 
$b_\mu$ according to 
\[
a_\mu+A_\mu=b_\mu+qA_\mu, 
\]
where the charge of the vortex $q$ equals 
\be
q=\frac{1}{2nk+1}. 
\label{q}
\ee
Rewriting the Lagrangian (\ref{L1}) we obtain the Lagrangian 
\be
L(b,A)=\nu_{0}\frac{1}{4\pi}\e^{\mu\nu\a}A_{\mu}\d_{\nu}A_\a+ 
\frac{1}{\a}\frac{1}{4\pi}\e^{\mu\nu\a}b_{\mu}\d_{\nu}b_\a+
\label{L2}
\ee
\[
V^{+}(\d-i{\bf b}-iq{\bf A}-i{\bf A}_1)^2V+ (b_0+qA_0)V^{+}V,  
\]
where the statistics parameter $\a$ (with respect to the Fermions) 
equals 
\be
\a=\frac{2k}{2nk+1}. 
\label{alpha}
\ee
We see from (\ref{L2}) that the field $V$ in fact represents the particles 
(anyons) with the charge $q$ and statistics $\a$. 
Differentiating the partition function corresponding to (\ref{L2}) over $A_0$ 
we obtain the additional part of the density $\rho_1=qn_q$, where $n_q$ is the 
density corresponding to the field $V$ (the density of Vortices). 
One could take the derivative over $A_0$ before integration over the field 
$\psi_0$. In this case we would obtain 
$\rho=\la\psi_0^{+}\psi_0+V^{+}V\ra$, where the integration measure includes 
the $\delta$- functions of the form (\ref{psi}). The density $\rho$ takes 
the form $\rho=\rho_0+2n\la \e_{ij}\d_ia_j\ra+\la V^{+}V\ra$, where according 
to eq.(\ref{L1}) $\e_{ij}\d_ia_j=-(\a/2)V^{+}V$. Thus we would obtain 
$\rho=\rho_0+\la V^{+}V\ra(-n\a+1)=\rho_0+q\la V^{+}V\ra$, which gives the 
same result $\rho=\rho_0+qn_q$.

Now one can easily calculate the FQHE filling fractions corresponding 
to the Jain-type hierarchy from the simple picture of anyons in the 
magnetic field. We perform the transformation of the type (\ref{2k}) 
with the integer parameter $k_1$ for the anyon wave function and demand 
that the integer number $n_1$ of the Landau levels in the effective 
magnetic field are completely filled. Thus we obtain the following equation 
for the density of Vortices $n_q$: 
\[
\frac{2\pi n_q}{H_1-2\pi n_q(\a-2k_1)}=n_1. 
\]
From this equation we obtain the following equation for the Anyon filling 
fraction $\nu_a= 2\pi n_q/H_1$: 
\[
\frac{1}{1/\nu_a-(\a-2k_1)}=n_1. 
\]
One can easily see that the total filling fraction $\nu$ equals  
\be
\nu=\nu_0+q^2\nu_a, 
\label{a}
\ee
where $q$, eq.(\ref{q}), is the charge of the Vortices. 
Substituting the values of $\a$, $q$, $\nu_a$ into the equation 
(\ref{a}) we obtain the following filling factors of Jain- type 
hierarchy:  
\be
\nu=\frac{n(2n_1k_1+1)+n_1}{(2nk+1)(2n_1k_1+1)+2kn_1},  
\label{nu}
\ee
where the numbers $n,k$ ($n_1,k_1$) are integers (positive or negative) 
with the allowed values given by the condition $\nu>0$. 
For $n=1$ ($m=2k+1$) the sequence (\ref{nu}) was first presented 
long time ago in ref.\cite{O1}.

Let us derive the same result directly from the Lagrangian (\ref{L1}) 
using the other method. This method does not rely on the values 
of the charge $q$, statistics $\a$, and the effective magnetic field 
$H_1$, and is based only on the general form of eq.(\ref{L1}).  
In fact we know that after the transformation (\ref{2k}) with the 
parameter $k_1$ integrating out the Vortex field $V$ in eq.(\ref{L1}) 
since $n_1$ Landau levels in the effective magnetic field are filled 
we obtain the effective Lagrangian of the form: 
\[
L_{eff}=n\frac{1}{4\pi}\e^{\mu\nu\a}(a+A)_{\mu}\d_{\nu}(a+A)_\a+ 
\frac{1}{2k}\frac{1}{4\pi}\e^{\mu\nu\a}a_{\mu}\d_{\nu}a_\a+
\]
\be
\frac{1}{2k_1}\frac{1}{4\pi}\e^{\mu\nu\a}c_{\mu}\d_{\nu}c_\a+ 
n_1\frac{1}{4\pi}\e^{\mu\nu\a}(a+A+c)_{\mu}\d_{\nu}(a+A+c)_\a. 
\label{L4}
\ee
Here the gauge field $c_{\mu}$ corresponds to the transformation (\ref{2k}) 
for the Vortex field. Integrating out the gauge fields $a_{\mu}$ and 
$c_{\mu}$ in eq.(\ref{L4}) we obtain the effective Lagrangian of the 
form $\nu\frac{1}{4\pi}\e^{\mu\nu\a}A_{\mu}\d_{\nu}A_\a$ where the 
conductivity (or the filling factor) $\nu$ is given exactly by the 
equation (\ref{nu}). Thus we see that under the certain assumptions 
the filling factor $\nu$ for the Jain-like hierarchy can be obtained 
even without the knowledge of the parameters $\a$, $q$, $H_1$ of the 
parent state. Presumably the Lagrangian (\ref{L4}) is analogous to 
the effective Lagrangian proposed in ref.\cite{LF03},\cite{LF98}.

\vspace{0.2in}

{\bf 3. Conclusion.} 

\vspace{0.2in}

In conclusion we proposed the effective Lagrangian which contains 
both the Fermi- and Bose- fields to describe the Jain-type hierarchical 
FQHE states. Using this effective Lagrangian (partition function) 
the charge of the quasiparticle excitations for the states of the Jain 
sequence was calculated. In contrast to the previous works 
\cite{LF03},\cite{LF98} the conductivity $\nu$ was calculated by means 
of considering the system of anyons in the external magnetic field. 
and the filling fraction was calculated by means of filling of the 
Landau levels in the effective magnetic field. 
The FQHE filling fractions are found from the condition of the 
non-zero coefficient of the Chern-Simons term in the effective 
Lagrangian. The comparison of the prediction (\ref{nu}) with the 
results of the experiments is given in ref.\cite{LF03}.

\end{document}